\documentclass[sigconf]{acmart}
\usepackage{multirow}
\usepackage{pifont}
\usepackage{makecell}
\usepackage{graphicx}
\usepackage{amsmath}
\AtBeginDocument{%
  }

\setcopyright{acmlicensed}
\copyrightyear{2025}
\acmYear{2025}
\setcopyright{acmlicensed}\acmConference[SIGIR '25]{Proceedings of the 48th International ACM SIGIR Conference on Research and Development in Information Retrieval}{July 13--18, 2025}{Padua, Italy}
\acmBooktitle{Proceedings of the 48th International ACM SIGIR Conference on Research and Development in Information Retrieval (SIGIR '25), July 13--18, 2025, Padua, Italy}
\acmDOI{10.1145/3726302.3729978}
\acmISBN{979-8-4007-1592-1/2025/07}

\begin{document}

\title{FIM: Frequency-Aware Multi-View Interest Modeling for Local-Life Service Recommendation}

\author{Guoquan Wang}
\authornote{Both authors contributed equally to this research.}

\orcid{0009-0001-7791-1203}
\author{Qiang Luo}
\authornotemark[1]

\orcid{0009-0001-7791-1203}
\affiliation{%
  \institution{Kuaishou Technology}
  \city{Beijing}
  \country{China}
}
\email{wangguoquan03@kuaishou.com}
\email{luoqiang@kuaishou.com}

\author{Weisong Hu}
\affiliation{%
  \institution{Kuaishou Technology}
  \city{Beijing}
  \country{China}}
\email{huweisong@kuaishou.com}

\author{Pengfei Yao}
\affiliation{%
  \institution{Kuaishou Technology}
  \city{Beijing}
  \country{China}}
\email{yaopengfei03@kuaishou.com}

\author{Wencong Zeng}
\affiliation{%
  \institution{Kuaishou Technology}
  \city{Beijing}
  \country{China}}
\email{zengwencong@kuaishou.com}

\author{Guorui Zhou}
\authornote{Corresponding author.}
\affiliation{%
  \institution{Kuaishou Technology}
  \city{Beijing}
  \country{China}}
\email{zhouguorui@kuaishou.com}

\author{Kun Gai}
\affiliation{%
  \institution{Unaffiliated}
  \city{Beijing}
  \country{China}}
\email{gai.kun@qq.com}

\renewcommand{\shortauthors}{Guoquan Wang et al.}
\begin{abstract}
People's daily lives involve numerous periodic behaviors, such as eating and traveling. Local-life platforms cater to these recurring needs by providing essential services tied to daily routines. Therefore, users' periodic intentions are reflected in their interactions with the platforms.
There are two main challenges in modeling users' periodic behaviors in the local-life service recommendation systems: 1) the diverse demands of users exhibit varying periodicities, which are difficult to distinguish as they are mixed in the behavior sequences; 2) the periodic behaviors of users are subject to dynamic changes due to factors such as holidays and promotional events.
Existing methods struggle to distinguish the periodicities of diverse demands and overlook the importance of dynamically capturing changes in users' periodic behaviors.
To this end, we employ a \textbf{F}requency-Aware Multi-View \textbf{I}nterest \textbf{M}odeling framework (\textbf{FIM}).
Specifically, we propose a multi-view search strategy that decomposes users' demands from different perspectives to separate their various periodic intentions. 
This allows the model to comprehensively extract their periodic features than category-searched-only methods.
Moreover, we propose a frequency-domain perception and evolution module. This module uses the Fourier Transform to convert users' temporal behaviors into the frequency domain, enabling the model to dynamically perceive their periodic features.
Extensive offline experiments demonstrate that FIM achieves significant improvements on public and industrial datasets, showing its capability to effectively model users' periodic intentions. 
Furthermore, the model has been deployed on the Kuaishou local-life service platform. Through online A/B experiments, the transaction volume has been significantly improved.
\end{abstract}




\maketitle

\section{Introduction}
By providing convenient services such as eating and traveling, local-life platforms (such as Kuaishou, Douyin, and Meituan) play an increasingly significant role in our daily lives~\cite{meituanlocal}.
Generally, these platforms propose to model users' interests by evaluating their historical interactions with platforms (such as clicking, liking, and purchasing)~\cite{sigir,icde}. 
They customize offerings in alignment with users' interests, thus increasing the consumers' propensity to purchase~\cite{foodlocal}.
However, there are two challenges that complicate the precise modeling of user interests in local-life platforms.

\begin{figure}[t]
\centerline{\includegraphics[width=0.47\textwidth]{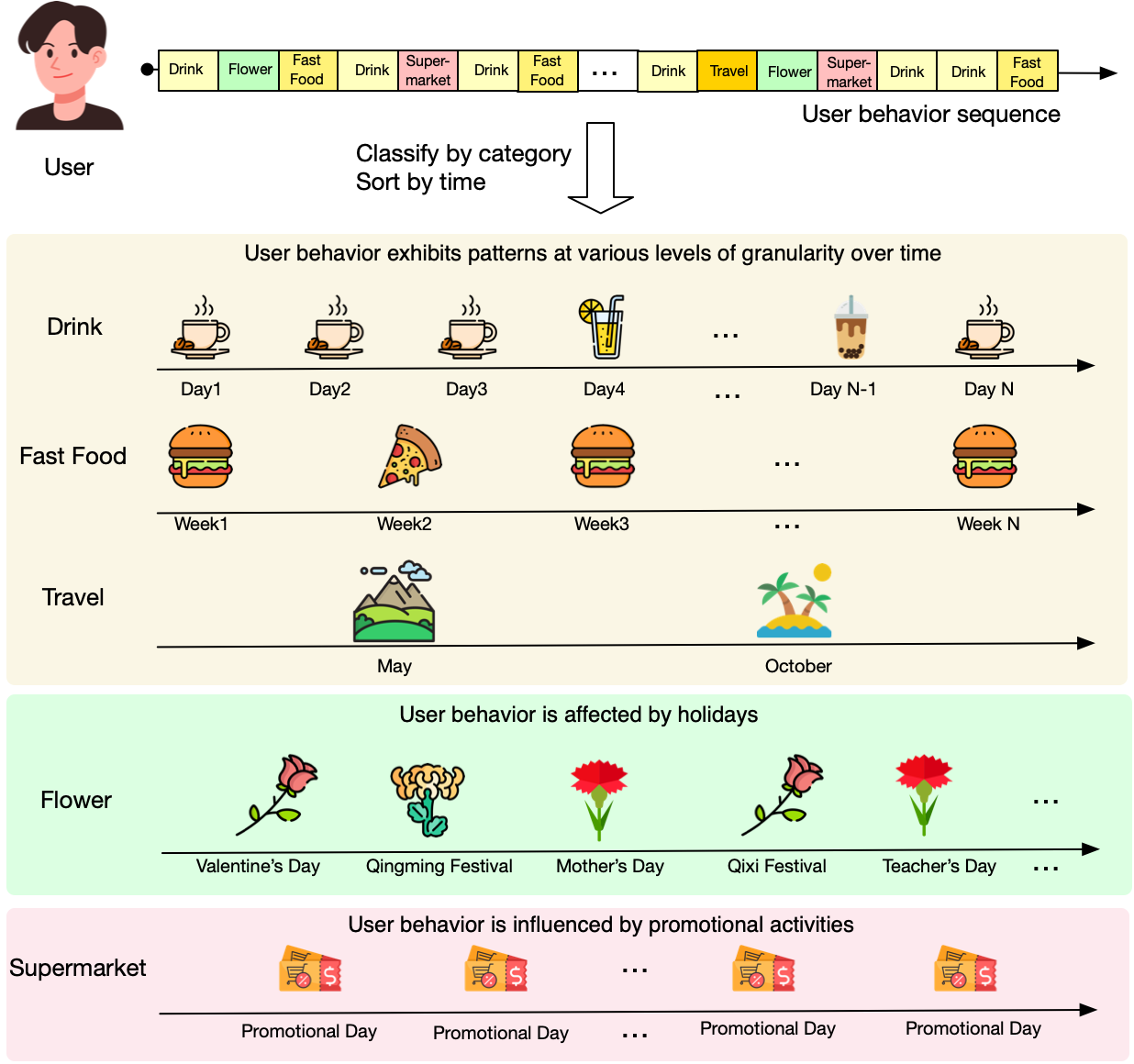}}
\centering
\caption{
Illustrates the motivation behind our research. The behavioral sequences of users regarding local-life goods exhibit an inherent periodicity. Additionally, certain holidays or promotional activities also influence the purchasing behavior of users.
}
\label{fig1}
\end{figure}

Firstly, in local-life scenarios, the varied purchasing needs of users exhibit distinct periodicities that are intricately blended into their behavioral sequences.
On one hand, users exhibit different consumption periods for various categories of products. As shown in the brown section of Figure \ref{fig1}, the user has a habit of purchasing drinks daily, whereas his travel frequency is quarterly.
On the other hand, purchasing decisions are influenced by multiple factors~\cite{cikmlocal}. For example, 
when selecting a restaurant, users typically consider its brand reputation, the pricing of dishes, and whether it aligns with specific dietary categories (e.g., vegetarian, gluten-free). These preferences vary across different time periods. 
Therefore, it is necessary to distinguish multiple demand preferences of users and their complex interactions regarding different item attributes. 

Secondly, the periodic behaviors of users are subject to dynamic changes due to factors such as holidays and promotional events.
Previous user interest modeling methods often emphasize modeling items that show up frequently in a user's past activity~\cite{twin,meituanlocal}. 
However, in local-life platforms, users exhibit different purchase intentions for the same item at various time granularities.
For example, promotional activities may lead users to repeatedly purchase the same type of item or stock up on group-buying vouchers shown in the pink part of Figure~\ref{fig1}.
During specific holidays, users might be influenced by short-term interests and purchase exploratory items, such as buying bouquets shown in the green part of Figure~\ref{fig1} for Mother's Day. 
These factors influence users' purchasing frequency, altering their inherent periodic consumption habits.




User interest modeling has garnered extensive attention from both academia and industry~\cite{hetelocal,foodlocal,meituanlocal,meituanlocal2}.
Early user interest modeling methods employed attention mechanisms to extract behavioral patterns for recommendation purposes~\cite{din,sasrec}. Recent approaches focus on modeling users' lifelong behaviors~\cite{eta,qin,twin}. For instance, SIM~\cite{sim} models lifelong sequential behaviors through two cascading stages. 
However, the aforementioned methods typically model users' non-periodic interests based on single-category information of their historical interaction sequences.
Conversely, local-life platforms face greater challenges due to the complex periodic relationships formed between users and various items through promotional activities~\cite{fajuan}. 
Therefore, it is important to make exploratory and repetitive modeling of local-life users’ periodic interests from multiple perspectives.

To this end, we propose a \textbf{F}requency-Aware Multi-View \textbf{I}nterest \textbf{M}odeling framework (\textbf{FIM}) for user interest modeling in local-life scenarios. 
Specifically, we introduce a multi-view search strategy (MSS) to uncover the periodic preferences of users' diverse demands.
MSS distinguishes users' needs from four different perspectives (price, brand, category, and author) to analyze user behavior patterns. In each perspective, it filters out subsequences unrelated to the user's demands to capture the potential periodic characteristics. MSS not only reveals changes in user preferences across different time periods but also efficiently models user interests.

Moreover, we propose a Frequency-domain Perception and Evolution Module (FPEM)  to capture users' repetitive and exploratory preferences. 
FPEM first transfers the multi-view subsequences from the time domain to the frequency domain. It then uses filters to divide them into different frequency bands for separate modeling, thereby capturing users' purchasing behaviors across different periods.
Subsequently, it employs a frequency-evolution module to capture the evolution of users' interests.
Unlike other methods, FPEM functions as a plug-and-play module that enhances recommendation performance across various user interest modeling methods. 
Additionally, it provides a pathway to apply frequency domain variations to better capture the periodic user interests in local-life service platforms.

In summary, our contributions are as follows:
\begin{itemize}
\item We emphasize the challenges of user interest modeling in local-life service scenarios, and further design FIM to address those challenges. We design MSS to separate users' periodic preferences across different needs and introduce FPEM to dynamically capture the evolution of their periodic intentions in the frequency domain.
\item A plug-and-play frequency-domain perception and evolution module is proposed to model users' periodic interests with dual repetitive and exploratory preferences. Extensive experiments demonstrate that this module can be added to other methods and improve accuracy.
\item We conduct extensive offline experiments on Kuaishou's 13 million-scale industrial dataset and other open-source datasets, and validate the effectiveness of FIM through online A/B testing. Our ablation studies confirmed our effectiveness, demonstrating that FIM offers significant online advantages.

\end{itemize}

\begin{figure*}[t]
\centerline{\includegraphics[width=1.0\textwidth]{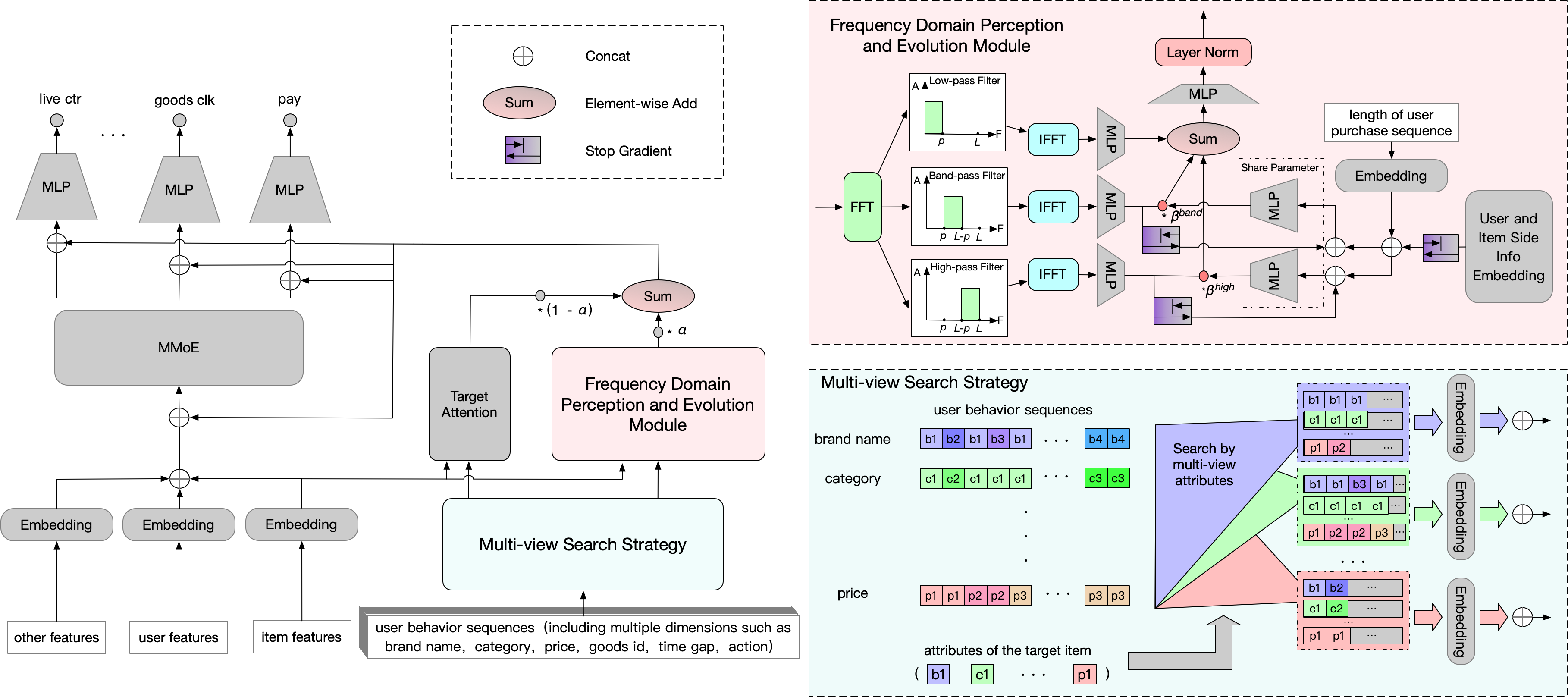}}
\centering 
\caption{The pipeline of the proposed FIM. It comprises three main components: a multi-view search module, a frequency-domain perception and evolution module, and a prediction module incorporating MMoE.}
\label{fig2}
\end{figure*}

\section{Related Work}

\textbf{User Interest Modeling} is the process of predicting user preferences by analyzing their historical behavior data. 
Early works employed memory networks to capture user interests~\cite{early1, early2}. In recent years, deep learning-based methods have achieved great success in this task~\cite{din,dien,caser}. 
DIN~\cite{din} captures users' varying interests in different candidate items by introducing target attention. 
DIEN~\cite{dien} builds on DIN by incorporating temporal relations in user historical behavior to capture the evolution of user interests. However, these methods fail to precisely capture user interests for a specific candidate item when the user behavior sequence is long.
To address this issue, SIM~\cite{sim} retrieves relevant subsequences from long historical sequences to model user behavior. 
Following this, ETA~\cite{eta} explores the use of locality-sensitive hashing methods and Hamming distance to retrieve relevant behaviors. 
TWIN~\cite{twin} achieves consistency in the two-stage life-long behavior modeling problem and accelerates the computation of target attention. 
QIN~\cite{qin} proposes two cascade units to filter raw user behaviors and reweight the behavior subsequences. 
Interest Clock~\cite{interestclock} encodes time-based behaviors into a 24-hour clock representation and then uses gaussian smoothing to aggregate nearby hours' preferences into a final embedding.
Unlike existing studies, we fully extract multi-view features of users in local life scenarios, allowing us to handle long user behavior sequences efficiently.

\textbf{Frequency Domain Learning} plays a pivotal role in the field of digital signal processing. In practice, the Fast Fourier Transform (FFT) has been widely used in computer vision~\cite{aaaifft20,aaaifft24} and natural language processing~\cite{fftnlp1,aaainlp2}. Due to its powerful capabilities, FFT has been widely used in the field of sequence recommendation. FMLP-Rec~\cite{filmlp} introduces the filtered MLP to remove noise in the frequency domain. FEARec~\cite{fearec} captures key features in user sequences through the auto-correlation mechanism implemented by FFT. Moreover, BSARec~\cite{bsarec} balances inductive bias and self-attention by applying the Fourier transform. Previous approaches have focused on denoising in the frequency domain to achieve precise recommendations. However, users in local-life scenarios often exhibit a certain degree of periodicity~\cite{meituanlocal}, which has not been adequately taken into account by the above approaches. To this end, we propose a frequency-domain perception and evolution module to capture the periodic interests of users in local-life scenarios.

\section{Preliminaries}
\subsection{Problem Definition}
\textbf{Notations.} In this work, the item set is given as $\mathcal{M}=\{m_i | i \in 1,2, \dots, |\mathcal{M}|\}$ and the user set is denoted by $\mathcal{U}=\{u_i | i \in 1, \dots, |\mathcal{U}|\}$, where $|\mathcal{M}|$ denotes the number of items and $|\mathcal{U}|$ denotes the number of users.
In this way, each chronologically ordered sequence is denoted by $S_u = [m_1^{(u)}, m_2^{(u)}, \dots, m_N^{(u)}]$, where $u \in \mathcal{U}$, and $m_i^{(u)} \in \mathcal{M}$ is the $i$-th item that user $u$ interacts with, and $N$ is the sequence length. $m_t^{(u)}$ is the target item that user $u$.
For $S_u$, it is embedded as $E_u = [e_1^{(u)}, e_2^{(u)}, \dots, e_N^{(u)}]$, where $e_i^{(u)}$ is the embedding of item $m_i^{(u)}$.
We can then formulate the problem of user interests for local-life recommendation systems as follows:

\noindent \textbf{Input:} The set of chronologically ordered sequences for all users: $S = \{S_1, \dots, S_{|\mathcal{U}|}\}$. The target item set for all users: $\{m_t^{(1)}, \dots, m_t^{|\mathcal{U}|}\}$.

\noindent \textbf{Output:} 
Probability of a user $u$ to purchase the target item $m_t^{(u)}$.


\subsection{Frequency Spectrum Representation}
The Fourier transform can decompose a chronologically ordered sequence of user behaviors into their constituent frequencies. This proves particularly advantageous for modeling user interests in local-life scenarios, as it facilitates the identification of periodic or exploratory patterns in the data. With this in mind, 
given the user sequence embedding matrix $E_u \in \mathbb{R}^{N \times D}$ and $E_u = \mathbf{x}$, we could execute Discrete Fourier Transform (DFT) denoted as $\mathcal{F}(\cdot)$ along the sequence dimension to transform the input item representation matrix $\mathbf{x}$ to the frequency domain:
\begin{equation}
\mathbf{X}[k]=\mathcal{F}\left(\mathbf{x}[i]\right)=\sum_{i=0}^{N-1} \mathbf{x}[i] e^{-j(2 \pi / N) k i}
\label{eq1}
\end{equation}
where $\mathbf{X}[k]$ is the spectrum of $\mathbf{x}[i]$ at the frequency $2\pi k/N$. $\mathcal{F}(\cdot)$ is the 1D DFT along the sequence dimension, and $j$ is the imaginary unit. Due to the conjugate symmetric property in the frequency domain~\cite{fearec}, only half of the spectrum is used in $\mathbf{X} \in \mathbb{C}^{L \times D}$. Thus, $L = \lceil {N}/{2} \rceil + 1$.
In this paper, we apply the Fast Fourier transform (FFT) \cite{fftorigin} to efficiently compute the DFT, reducing time complexity from $O(n^2)$ to $O(n \log n)$. The inverse Fast Fourier transform (IFFT) works similarly.

\section{Methodology}
\subsection{Overview}
The overview of our proposed FIM is depicted in
Figure~\ref{fig2}. Firstly, the Multi-view Search Strategy (MSS) conducts searches across various attributes of the user's behavioral sequences to extract subsequences that are implicitly periodic and related to specific items. Then, the Frequency domain Perception and Evolution Module (FPEM) employs the FFT to transition these subsequences from the time domain to the frequency domain. After passing through a series of filtering operations tailored to different frequency bands and then inverse transforming back to the time domain, it achieves perception in the frequency domain. Subsequently, it integrates the filtered outputs in a user-personalized manner to evolve the user's behavioral cycles.
Finally, the integrated outcomes combine multi-view time-domain attention results with frequency-aware results before passing through a prediction layer for user interest modeling.

\subsection{Multi-View Interest Modeling}

To accurately model user interests, it is essential to analyze their long-term behavior to understand their consumption preferences, transitions, and motivations. Unfortunately, existing public datasets either have user behavior sequences that are too short or lack sufficient historical behavior data. For instance, in the widely used Amazon dataset~\cite{amazondata}, the average historical sequence length for each user does not exceed 200. In the Taobao dataset~\cite{taobaodata}, there are no more than five types of user behaviors. Therefore, we collect an industrial dataset from the Chinese short-video platform Kuaishou to model user interests in the context of local life scenarios.

In our Kuaishou industrial dataset, the user behavior sequence encompasses nine dimensions: goods ID, author ID, source domain, action, brand name, product category, time span, product price, and payment amount, to comprehensively cover user interest expressions across multiple dimensions. 
Therefore, each view contains nine different user attributes, i.e.,  $|\mathcal{A}|$ = 9, with the meaning of the number of attributes for different views. 

The behavior sequences of local-life users are often exceptionally long, leading to significant resource consumption when directly using them for modeling user interests~\cite{dien}. It is generally unacceptable in practical applications. 
It is generally unacceptable in practical applications.
In fact, only a portion of the user sequences are crucial for interest modeling, making the selection of relevant user behaviors essential. 
However, the existing search strategy only considers filtering based on categories~\cite{sim,qin}.
The purchasing intentions of local-life users are influenced by various factors such as price, category, and brand~\cite{cikmlocal}. Inspired by it, we propose a multi-view search strategy to reduce the number of user behavioral sequences for interest modeling. 
Specifically, we match the entire user behavior sequence from four different perspectives: the author ID of the target product, brand name, product category, and product price, i.e., $|\mathcal{M}|$ = 4, with the meaning of the number of views for different search strategies.
Mathematically, the Top-K relevant behavior sequences with score $r_i$ are selected as subsequences:
\begin{equation}
r_i = \left\{
\begin{array}{ll}
\text{sign}(m^{v}_{i}, m^{v}_{t}) & \text{hard-search} \\[0.5em]
(W_b e^{v}_{i} \odot W_t e^{v}_{t}) & \text{soft-search}
\end{array}
\right.
\label{eq2}
\end{equation}
where $m_{i}^{v}$ and $m_{t}^{v}$ in the hard-search strategy denote the $v$-th view of $i$-th item and target item correspondingly.
$e_i^v$ and $e_t^v$ in the soft-search strategy represent the concatenation of vectors $e_i^{v,a}$ over all attributes $a$ in set $\mathcal{A}$, where $i$ and $t$ denote $i$-th item and target item, respectively. $W_b$ and $W_t$ are the parameters of weight. 

After the search process, the multi-attribute behavior sequence of different views serves as the key/value in the target attention structure~\cite{din}, while the candidate item features act as the query. Finally, the outputs of target attention from the four views are concatenated to serve as the final output of the multi-view interest modeling module.

\subsection{Frequency-Aware Interest Modeling}
In local-life scenarios, the same user may exhibit different periodic behavior patterns, such as repetitive, seasonal, or exploratory purchases. When observing user behavior in the time domain, some periodic trends may be obscured by noise or data complexity. To this end, we transform the data into the frequency domain to perceive users' underlying purchasing cycles and preferences. Mathematically, we follow Eq.~\ref{eq1} to calculate the frequency representation $F_u \in \mathbb{C}^{L \times D}$ of the user embedding sequence $E_u \in \mathbb{R}^{N \times D}$ as below:
\begin{equation}
F_u = \mathcal{F}(E_u)
\label{eq3}
\end{equation}
where $\mathcal{F}(\cdot)$ is the conjugate symmetric fast fourier transform. In light of this, we can extract the desired frequency components through different filters as follows:
\begin{equation}
\text{filter}(c) = \left\{\begin{matrix}1, & c\in [c_b, c_e) \\0,
  &\text{otherwise}
\end{matrix}\right.
\label{eq4}
\end{equation}
where $c_b$, $c_e$ refers to the starting and ending frequency of the filter, respectively.

By varying the values of $c_b$ and $c_e$, we obtained three distinct types of filters $F_u^{low},F_u^{band},F_u^{high}$ to observe user behavior from different frequency perspectives. When $c_b$ is taken to be 0 and $c_e$ is taken to be $p$, the filter can be considered as a low-pass filter, where $p$ is the hyperparameter that controls the range of the low-pass filter. The low-pass filter helps to preserve the slow changing trends in the sequence of user behavior and filters out those rapidly changing fluctuations. It infers the overall user behavior trend by analyzing the long-term shopping habits and repetitive interests of the user.
When $c_b$ is taken as $L-p$ and $c_e$ is taken as $L$, the filter becomes a high-pass filter, where $L$ represents the length of the entire frequency band, which can be calculated by Eq.~\ref{eq1}. The high-pass filter retains the significantly changing part of the user's interest, reflecting the user's episodic shopping behavior, which is manifested in the user's sudden emergence of points of interest and consumption impulses. This is suitable for analyzing short-term promotions in local-life service environments.
When $c_b$ is taken to be $p$ and $c_e$ is taken to be $L-p$, the filter becomes a band-pass filter, which is able to filter out either too slow or too fast changes at the same time and capture the characteristics of the intermediate frequencies of user behavior.

To achieve a more comprehensive understanding of user behavior, we employ a frequency-aware mechanism that bridges the three filters, enabling more precise modeling of the evolution in user behaviors.
We first apply the Inverse Fast Fourier Transform (IFFT) to the input samples 
, extracting features from three different frequency bands using filters:
\begin{equation}
f_u^{low},f_u^{band},f_u^{high} = \mathcal{F}^{-1}(F_u^{low},F_u^{band},F_u^{high}).
\label{eq5}
\end{equation}

Since there are noisy behaviors of the user in the high and mid-frequencies, we propose to weaken the influence of the high and mid-frequency bands on the final discrimination to some extent. Therefore we design a frequency-aware module to suppress the user's behavior, which can be defined as:
\begin{align}
\beta^{band}  &= \sigma (W_2\phi(W_1X^{band}+B_1)+B_2), \\
\beta^{high}  &= \sigma (W_2\phi(W_1X^{high}+B_1)+B_2) 
\end{align}
\label{eq6}
where $X^{band}$ denotes the feature after concatenation of $I_u$ and $f_u^{band}$. $I_u$ represents the embedding set of user and target item side info (e.g., user's age, gender; item's brand, price) as well as the length of the user's purchase sequences. 
We retain only the gradient of the user's purchase sequence length in Eq.~\ref{eq6} for backpropagation to avoid mutual interference between representation learning.
$W_1$ and $W_2$ represent the weights of the features, and $B_1$ and $B_2$ correspond to the biases of the features. $\phi(\cdot)$ and $\sigma(\cdot)$ represent the ReLU and sigmoid activation functions, respectively. The irrelevant components in the high and mid-frequency bands are finally restrained by the following fusion process:
\begin{align}
f^{fuse}_{u} &= f^{low}_u +\beta^{band}f^{band}_u+\beta^{high}f^{high}_u, \\
f_{u}  &= \text{LayerNorm}(f_u^{fuse} + E_u).
\label{eq7}
\end{align}

The input tensor has dimensions $I \times D$, where $I$ is the sequence length and $D$ is the feature dimension. Existing methods such as FEARec \cite{fearec} and BSARec \cite{aaaifft24} have a time complexity for self-attention on the $I$ dimension as $O(I^2 \times D)$ and for MLP on the $D$ dimension as $O(I \times D^2)$, resulting in a total time complexity of $O(I^2 \times D + I \times D^2)$.
Our proposed method, FPEM, involves performing a Fast Fourier Transform on the $I$ dimension with a time complexity of $O(I \log I)$ and computing MLP on the $D$ dimension with a time complexity of $O(I \times D^2)$, giving a total time complexity of $O(I \log I + I \times D^2)$, which is lower than other methods.

\subsection{Prediction and Optimization}
Due to the varying number of behaviors among different users, it is necessary to perform average pooling over the time dimension on the output feature $H_u$ of the multi-view interest modeling network and the output feature $f_u$ of the frequency-aware modeling network. The two averaged features $\bar{H}_u$ and $\bar{f}_u$ are then combined through weighted fusion to obtain the overall vector:
\begin{equation}
Z_u = (1-\alpha) \bar{H}_u + \alpha\bar{f}_u
\label{eq8}
\end{equation}
where $\alpha$ represents the coefficient used to balance the importance of the two modules.
The final vector is fed into the Multi-gate Mixture-of-Experts (MMoE)~\cite{mmoe} for multi-task learning, which allows the model to share representations across tasks while maintaining task-specific expertise. Additionally, a residual connection is applied to the output of the MMoE to enhance the model's ability to capture complex patterns and mitigate the risk of vanishing gradients. The final output is then processed through multiple distinct classification heads, such as click-through rate prediction, watch time estimation, and purchase rate prediction, to tailor the model's predictions to the specific requirements of each task.

For optimization, we employ the Binary Cross-Entropy 
 (BCE) loss as the loss function for our model. Its mathematical formulation is as follows:
\begin{equation}
\mathcal{L} = -\left(y \log(\hat{y}) + (1-y) \log(1-\hat{y})\right)
\label{eq9}
\end{equation}
where $y$ denotes the ground truth (0 or 1) and $\hat{y}$ represents the predicted probability (ranging from 0 to 1).

\section{Experiment}


\begin{table*}[!h]
\centering
\caption{Performance comparison of different methods on public and industrial datasets. The best results are in boldface and the second-best results are underlined. ‘Improv.’ indicates the relative improvement against the best baseline performance. Methods marked with * use a search strategy, while the remaining methods directly model on the complete sequence. 
}
\renewcommand{\arraystretch}{1.5}
\setlength\tabcolsep{4.2pt}
\resizebox{\textwidth}{!}{%
\begin{tabular}{c|c|ccccc|ccccc|ccc}
\toprule
\textbf{Dataset} & \textbf{Metric} & \textbf{DIN} & \textbf{DIEN} & \textbf{SASRec} & \textbf{Caser} & \textbf{BSARec} & \textbf{SIM-Hard*} & \textbf{SIM-Soft*} & \textbf{ETA*} & \textbf{QIN*} & \textbf{TWIN*} & \textbf{FIM-Hard*} & \textbf{FIM-Soft*} & \textbf{Improv.}\\  \cmidrule(r){1-15}
\multirow{2}{*}{Book Dataset} & AUC                     & 0.7331 & 0.7411 & 0.7324  & 0.7370 & 0.7325 & 0.7331    & 0.7485    & 0.7443 & \underline{0.7494} & 0.7479 & \(0.7544 \pm 0.0123\) & \(\textbf{0.7583} \pm 0.0015\) & 1.19\%  \\
                         & GAUC                    & 0.7108 & 0.7146 & 0.7128  & 0.7109 & 0.7038 & 0.7069    & 0.7258    & 0.7255 & \underline{0.7278} & 0.7259 & \(0.7279 \pm 0.0093\)  & \(\textbf{0.7341} \pm 0.0012\) & 0.87\%  \\ \cmidrule(r){1-15}
\multirow{2}{*}{Taobao Dataset} & AUC                    & 0.7937 & 0.8010 & 0.7981  & 0.7946 & 0.7963 & 0.7925    & 0.7987    & 0.8009 & 0.8036 & \underline{0.8056} & \(0.8060 \pm 0.0343\) & \(\textbf{0.8130} \pm 0.0288\) & 0.92\%  \\
                         & GAUC                    & 0.6682 & 0.6805 & 0.6745  & 0.6688 & 0.6767 & 0.6683    & 0.6708    & 0.6815 & \underline{0.6884} & 0.6876 & \(0.6944 \pm 0.0135\) & \(\textbf{0.7004} \pm 0.0162\) & 1.74\%\\ \cmidrule(r){1-15}
\multirow{2}{*}{Kuaishou Dataset} & AUC                    & 0.8888    & 0.8820    & 0.8900 & - & 0.8927 & \underline{0.8930}  & 0.8840  & 0.8844 & 0.8879 & 0.8862 & \(\textbf{0.9007} \pm 0.0004\) & \(0.8875 \pm 0.0018\) & 0.86\% \\
                         & GAUC                    & 0.6398    & 0.6405    & 0.6363  & - & 0.6487 & \underline{0.6698}  & 0.6487 & 0.6421 & 0.6476 & 0.6467 & \(\textbf{0.6944} \pm 0.0005\) & \(0.6607 \pm 0.0021\) & 3.67\%\\ \bottomrule
\end{tabular}%
}
\label{baseline}
\end{table*}

In this section, we briefly describe the experimental settings and evaluate our proposed model through extensive experiments to address the following research questions:
\begin{itemize}
    \item \textbf{RQ1:} How does FIM perform compared with state-of-the-art recommendation models?
    \item \textbf{RQ2:} What is the effect of MSS and FPEM modeling in our proposed FIM?
    \item \textbf{RQ3:} How does FIM perform in the real-world online recommendation scenarios with practical metrics?
\end{itemize}

\subsection{Experimental Settings}

\textbf{Datasets.} Model comparisons are conducted on two public datasets:  Taobao Dataset \cite{taobaodata,taobaodata2} and Amazon Dataset \cite{amazondata},  as well as an industrial dataset collected from a prominent Chinese short-video platform, Kuaishou.
\begin{itemize}
    \item \textbf{Taobao} \cite{taobaodata,taobaodata2}: Data is collected from Taobao, which is China's leading e-commerce platform. The dataset includes user actions such as clicks, adding items to the cart, likes, and purchases during the period from November 25th to December 3rd, 2017. In our evaluation, we categorize user clicks as negative interactions, while all other actions are considered positive. The training data consists of all instances up to December 1st, with the final day used as the test set.
    \item \textbf{Amazon (Book) }\cite{amazondata}: This dataset is sourced from Amazon and consists of product details such as titles, authors, ratings, and reviews. We divide the latest 10 user behaviors into short-term features and the previous 90 into long-term features. The preprocessing approach applied to this dataset is consistent with that used for SIM \cite{sim}.
    \item \textbf{Kuaishou}: This dataset is sourced from a prominent Chinese short-video platform, Kuaishou. We extract a portion of data logs spanning a month, specifically from April 21st to May 20th, 2024, by randomly selecting 10\% of users. The initial 23 days' data is allocated for training purposes and the remaining 7 days for testing.
\end{itemize}

\textbf{Evaluation Metrics.} For offline evaluation, we adopt two widely used metrics to measure predictive accuracy: AUC and GAUC. AUC signifies the probability that a positive sample’s score is higher than that of a negative one, reflecting a model’s ranking ability. GAUC performs a weighted average over all user’s AUC, which alleviates the negative impact of unbalanced distributions across users. 
For online A/B tests, we apply Gross Merchandise Value (GMV) to indicate a total sales monetary-value for merchandise sold over a certain period of time.

\textbf{Implementation Details.} All models are implemented using TensorFlow \cite{tensorflow}. The Adam optimizer \cite{adam} is employed with a learning rate of 0.01. The batch size are set to be 256, 256, and 6144 for Taobao, Book and Kuaishou dataset, respectively. The dimension of the item embeddings is set to 4 for Taobao and Book datasets, and 16 for Kuaishou dataset. $\alpha$ is set to 0.5.

\textbf{Baselines.} For lifelong interest modeling, we include SIM \cite{sim}, TWIN \cite{twin}, QIN \cite{qin}, and ETA \cite{eta}. For short-term interest modeling, we compare with Caser \cite{caser}, DIN \cite{din}, DIEN \cite{dien} and SASRec \cite{sasrec}. Additionally, we include BSARec \cite{bsarec}, the state-of-the-art model for frequency domain sequential recommendation, to validate the effectiveness of our FIM. The following is a detailed description of these methods.
\begin{itemize}
\item DIN \cite{din}: This method proposes an attention network to obtain the similarity between historical items and the target item, to calculate the interaction probability.
\item DIEN \cite{dien}: This method extends DIN by combining a recurrent neural network.
\item Caser \cite{caser}: This method adopts convolutional filters to extract the sequential patterns in user behaviors.
\item SIM \cite{sim}: This method adopts cascading search unit General Search Unit (GSU) and Exact Search Unit (ESU) to extract the relevant behaviors of the candidate item and applies target attention to model users’ interest. 
\item TWIN \cite{twin}: Two-Stage Interest Network adopts the same relevance metric between the target behavior and historical behaviors as the target attention in two cascading stages GSU and ESU.
\item QIN \cite{qin}: This method utilizes a query-dominant user interest network with cascade units to filter and reweigh user behavior subsequences.
\item ETA \cite{eta}:  This method employs a locality-sensitive hash-based efficient target attention network for end-to-end user behavior retrieval.
\item SASRec \cite{sasrec}: This method employs self-attention mechanisms to model user behavior sequences and to capture sequential preferences as context vectors.
\item BSARec \cite{bsarec}: This method leverages the Fourier transform to address the oversmoothing problem in Transformer-based sequential recommendation models by integrating low and high-frequency information. 
\end{itemize}

\subsection{Offline Evaluation (RQ1)}

As illustrated in Table \ref{baseline}, we compare our method with the state-of-the-art methods on both public datasets (Book and Taobao) and an industrial dataset (Kuaishou). On the public Book datasets, the FIM-Soft method achieves a relative improvement of 1.19\% and 1.39\% in AUC compared to QIN~\cite{qin} and TWIN~\cite{twin}, which model long sequential features. This enhancement is attributed to FPEM's effective capture of user's periodic interests. 
Embedding learning of user features becomes more challenging on industrial datasets, leading to sub-optimal results for methods using soft search strategies~\cite{sim,eta,qin,twin}. 
There are mainly two reasons for this. For one: hard-search conducts experiments under four different views on the Kuaishou dataset. While on the Book and Taobao datasets, it only performs hard-search on the category of items. For another, Kuaishou dataset has a larger candidate and user space than other datasets, so soft learning is more difficult to converge.
Nevertheless, our FIM-Hard approach with a multi-view hard search strategy surpasses the SIM-Hard~\cite{sim} model with category hard search, achieving relative improvements of 0.86\% in AUC and 3.67\% in GAUC.
To further elaborate on the variance and standard deviation of each experimental indicator, for the Book dataset, the variance of FIM-Soft’s AUC is 0.0015, and the variance of GAUC is 0.0012. For the Taobao dataset, the variance of FIM-Soft’s AUC is 0.0288, and the variance of GAUC is 0.0162. These variances indicate a relatively high level of model stability across different datasets, with lower variances suggesting more consistent performance. The consistently low variances, particularly in the Book dataset, highlight the robustness and reliability of the FIM-Soft model.

\subsection{Ablation Study (RQ2)}

\begin{table}[]
\centering
\caption{Ablation study of different components on Kuaishou dataset. ‘Time Improv.’ indicates the relative time improvement against the baseline performance. Methods marked with \# use a mulit-view interest search strategy, while the remaining methods directly model on the complete sequence. 
}
\renewcommand{\arraystretch}{1.5}
\setlength\tabcolsep{4.2pt}
\resizebox{1.0\columnwidth}{!}{%
\begin{tabular}{ccccc|cc|cc}
\toprule
\textbf{Author$^{\#}$} & \textbf{Brand$^{\#}$} & \textbf{Category$^{\#}$} & \textbf{Price$^{\#}$} & \textbf{FPEM} & \textbf{Param. (M)} & \textbf{Time Improv. (\%)} & \textbf{AUC(\%)} & \textbf{GAUC(\%)}\\
\midrule
\ding{55}  & \ding{55} & \ding{55} & \ding{55} & \ding{55} & 35.41 & - & 88.58 & 64.70 \\
\ding{55}  & \ding{55} & \ding{55} & \ding{55} & \checkmark  & 35.60 & 0.91 & 89.50 & 67.41 \\
\checkmark  & \ding{55} & \ding{55} & \ding{55} & \ding{55}  & 36.38 & 2.35 & 88.91 & 64.24 \\
\ding{55}  & \checkmark & \ding{55} & \ding{55} & \ding{55}  & 36.38 & 2.51 & 89.50 & 67.41 \\
\ding{55} & \ding{55} & \checkmark & \ding{55} & \ding{55} & 36.38 & 2.37 & 89.34 & 66.69 \\
\ding{55} & \ding{55} & \ding{55} & \checkmark & \ding{55} & 36.38 & 2.28 & 89.57 & 65.95 \\
\checkmark & \checkmark & \checkmark & \checkmark & \ding{55} & 36.76 & 5.25 & 90.01 & 67.54 \\
\checkmark & \checkmark & \checkmark & \checkmark & \checkmark & 36.93 & 6.37 & 90.07 & 69.44 \\
\bottomrule
\end{tabular}
}
\label{aball}
\end{table}

We conduct extensive ablation studies to assess the effectiveness of our FIM and its core modules. 
For the efficiency concern, we have presented detailed complexity analysis in Table \ref{aball}. This includes an examination of how the model's capacity, such as parameter size, is utilized and how it impacts the overall performance.
Regarding the evaluation, we have conducted in-depth experiments and presented the results in the tables. These results clearly show the performance gains and provide insights into how these improvements are achieved.
Specifically, we perform: 1) tests of various choices within the MSS module to confirm the effectiveness of the strategy; 
2) experiments on FPEM across different methods to validate the plug-and-play capability of the module; 
and 3) evaluations of different implementation strategies within the FPEM module to assess the rationality of the module's design.

\begin{figure}
    \centering
    \includegraphics[width=\columnwidth]{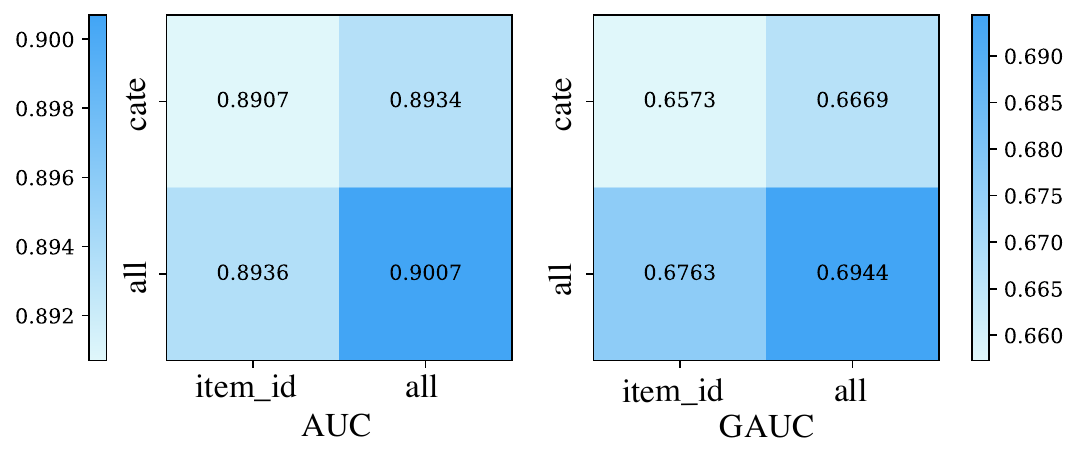}
    \caption{The ablation study of the MSS strategy on Kuaishou dataset, the vertical axis represents the matching views, and the horizontal axis represents the item attributes used.}
    \label{mss}
\end{figure}

\textbf{MSS Strategy Effectiveness Testing}. 
As shown in Figure \ref{mss}, we analyze the experimental results of our model on the industrial dataset when selecting different views and attributes. 
It can be seen that the strategy of only matching the category and using only the item\_id of the item, which is the hard search strategy of SIM, performs the worst. 
It can be seen that the SIM-Hard search strategy \cite{sim}, which conducts the category-based search on users' item\_id attribute, performs the worst.
The multi-view strategy can uncover more items related to user interests, and adding different matching views can bring a 0.33\% and 2.89\% relative improvement in AUC and GAUC, respectively. 
Moreover, when adding different attributes for representation learning, it can result in a 0.30\% and 1.46\% relative improvement in AUC and GAUC, respectively.
It can been found that multiple attributes provide the model with different perspectives to explore users' periodic features. When using both multi-view and multi-attribute approaches, i.e., MSS, the AUC and GAUC are improved by 1.12\% and 5.64\%, respectively, compared to the original strategy.
When analyzing the data in Table \ref{aball}, we can find that the gains from matching user behavior sequences through the Author perspective are relatively limited. This is because the same author may sell a variety of different products, and not all of them may meet the preferences of customers, so relying solely on author information may not be enough to accurately predict user behavior. In contrast, matching prices alone can bring significant gains because prices reflect customers’ spending power and purchasing levels to a certain extent, which is particularly important for predicting user behavior on local life platforms.
When all four views are used, the model achieves the highest AUC and GAUC values of 90.07\% and 69.44\% respectively, while the time consumption is improved by 6.37\%. This shows that integrating multiple perspectives can effectively improve the predictive performance of the model, especially the introduction of price information plays a key role in improving model performance. Therefore, when matching user behavior sequences, rationally selecting and combining multiple perspectives is an effective strategy to optimize model performance. By combining multiple perspectives, the model can more comprehensively capture user preferences and behavior patterns, thereby improving the accuracy and stability of predictions.

\begin{table}[]
\centering
\caption{Comparison of FPEM module performance across different soft search-based methods.}
\renewcommand{\arraystretch}{1.5}
\setlength\tabcolsep{4.2pt}
\resizebox{\columnwidth}{!}{%
\begin{tabular}{c|ccccccc}
\toprule
\multirow{2}{*}{\textbf{Method}} & \multicolumn{2}{c}{\textbf{Book Dataset}} & \multicolumn{2}{c}{\textbf{Taobao Dataset}} & \multicolumn{2}{c}{\textbf{Kuaishou Dataset}} & \multirow{2}{*}{\makecell{\textbf{Infer}\\\textbf{Speed}}} \\ \cmidrule(r){2-3} \cmidrule(r){4-5} \cmidrule(r){6-7}
         & \textbf{AUC}   & \textbf{GAUC}  & \textbf{AUC}   & \textbf{GAUC}  & \textbf{AUC} & \textbf{GAUC} &  \\ \cmidrule(r){1-8}
DIN      & 0.7331 & 0.7108 & 0.7937 & 0.6682 &  0.8889   & 0.6398     &   - \\
DIN+FPEM  
& +0.0065 & +0.0053 & +0.0048 & +0.0042 & +0.0045 & +0.0167

& +1.73\% \\ \cmidrule(r){1-8}
SIM      & 0.7485 & 0.7258 & 0.7987 & 0.6708 & 0.8840    & 0.6487     & - \\
SIM+FPEM  
& +0.0042 & +0.0044 & +0.0037 & +0.0037 & +0.0056 & +0.0056

&  +2.91\% \\ \cmidrule(r){1-8}
ETA      & 0.7443 & 0.7255 & 0.8009 & 0.6815 & 0.8844    & 0.6421     & - \\
ETA+FPEM  
& +0.0047 & +0.0034 & +0.0028 & +0.0018 & +0.0032 & +0.0094
  
& +0.54\% \\ \cmidrule(r){1-8}
QIN      & 0.7494 & 0.7278 & 0.8036 & 0.6884 & 0.8879    & 0.6477     & - \\
QIN+FPEM  
& +0.0034 & +0.0032 & +0.0035 & +0.0054 & +0.0025 & +0.0155

& +2.95\% \\ \cmidrule(r){1-8}
TWIN     & 0.7479 & 0.7259 & 0.8056 & 0.6876 & 0.8862    & 0.6467     & - \\
TWIN+FPEM 

& +0.0044 & +0.0046 & +0.0031 & +0.0031 & +0.0015 & +0.0148
&  +0.74\% \\ \cmidrule(r){1-8}
FIM (Ours)     & 0.7583 & 0.7341 & 0.8130 & 0.7004 & 0.9007    & 0.6944     &  - \\ \bottomrule
\end{tabular}%
}
\label{fft_plugin}
\end{table}

\textbf{FPEM Plug-and-Play Validation}. 
Table \ref{fft_plugin} illustrates the impact of the plug-and-play module on five different methods. 
Table \ref{fft_plugin} illustrates the impact of the plug-and-play module across various soft search-based methods, highlighting its impact on three different datasets: Book, Taobao, and Kuaishou dataset. It demonstrates that the integration of the FPEM module consistently enhances both AUC and GAUC metrics across all methods and datasets, indicating its robustness and effectiveness.
For instance, the SIM method benefits from the FPEM module with an AUC increase of 0.42\% on the Book dataset and a GAUC improvement of 0.56\% on the Kuaishou dataset. The inference speed for SIM also sees a notable enhancement of 2.91\%, indicating that the FPEM module not only improves predictive performance but also optimizes computational efficiency.
Similarly, TWIN \cite{twin} experiences a 0.44\% increase in AUC on Book dataset, while DIN \cite{din} sees a 0.42\% improvement in GAUC on Taobao dataset. This performance enhancement is consistently observed across all tested methods. Moreover, despite the performance gains, the module's effect on inference speed is negligible. For instance, the inference time for DIN increases by only 1.73\%, while the inference times for ETA \cite{eta} and TWIN remain virtually unchanged. This indicates that the module effectively enhances model performance without increasing inference time.
Finally, the our FIM method achieves the highest AUC and GAUC values across all datasets, with an AUC of 75.83\% and a GAUC of 73.41\% on the Book dataset, and an AUC of 90.07\% and a GAUC of 69.44\% on the Kuaishou dataset. This underscores the superior performance of the FIM method when integrated with the FPEM module.

\begin{table}[]
\centering
\caption{Ablation study of different implementations of FPEM. The first and second columns represent the methods for separating high and low frequencies, while the third and fourth columns represent the methods for fusing information from different frequency domains.}
\renewcommand{\arraystretch}{1.5}
\setlength\tabcolsep{4.2pt}
\resizebox{\columnwidth}{!}{%
\begin{tabular}{cc|cccccc}
\toprule
\multirow{2}{*}{\textbf{Trunct}} &
  \multirow{2}{*}{\textbf{Filter}} &
  \multirow{2}{*}{\textbf{$\beta$ Fusion}} &
  \multirow{2}{*}{\makecell{\textbf{Direct}\\\textbf{Fusion}}} &
  \multicolumn{2}{c}{\textbf{Book Dataset}} &
  \multicolumn{2}{c}{\textbf{Taobao Dataset}} \\ \cmidrule(r){5-6} \cmidrule(r){7-8}
  &   &   &   & \textbf{AUC}   & \textbf{GAUC}  & \textbf{AUC}   & \textbf{GAUC}  \\ \cmidrule(r){1-8}
\checkmark &  \ding{55}  & \checkmark &  \ding{55}  & 0.7583 & 0.7341 & 0.8130 & 0.7004 \\
\checkmark &  \ding{55}  &  \ding{55}  & \checkmark & 0.7389 & 0.7169 & 0.8075 & 0.6954 \\
 \ding{55} & \checkmark & \checkmark &  \ding{55}  & 0.7454 & 0.7238 & 0.8126 & 0.6982 \\
 \ding{55}  & \checkmark &  \ding{55}  & \checkmark & 0.7438 & 0.7099 & 0.8036 & 0.6903 \\ \bottomrule
\end{tabular}
}
\label{fft_design}
\end{table}

\begin{figure*}[!t]
    \centering
    \includegraphics[width=\textwidth]{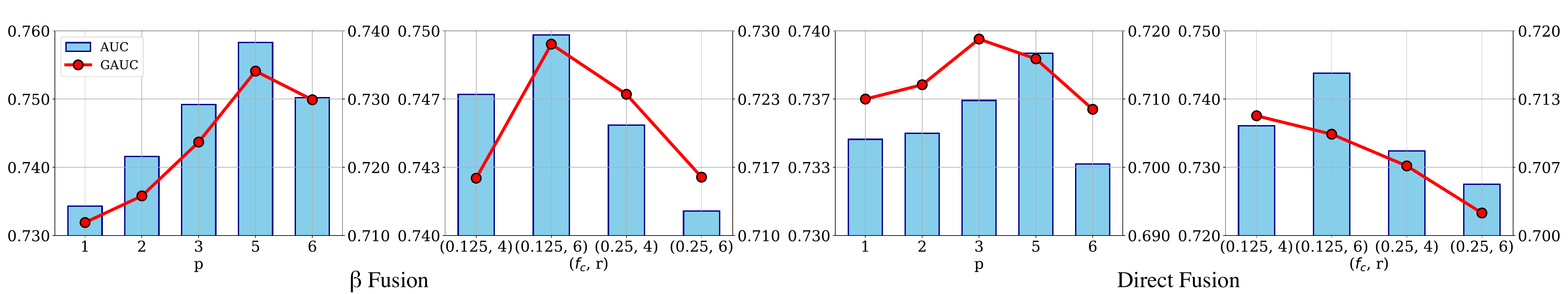}
 \caption{Comparison of truncation positions and filter parameters on FPEM's AUC/GAUC metrics for the Book dataset under various fusion strategies. $p$ is the truncation position, and $(f_c, r)$ represents the filter's cutoff frequency and order.}
    \label{c and fc}
\end{figure*}

\begin{figure*}[!t]
    \centering
    \includegraphics[width=\textwidth]{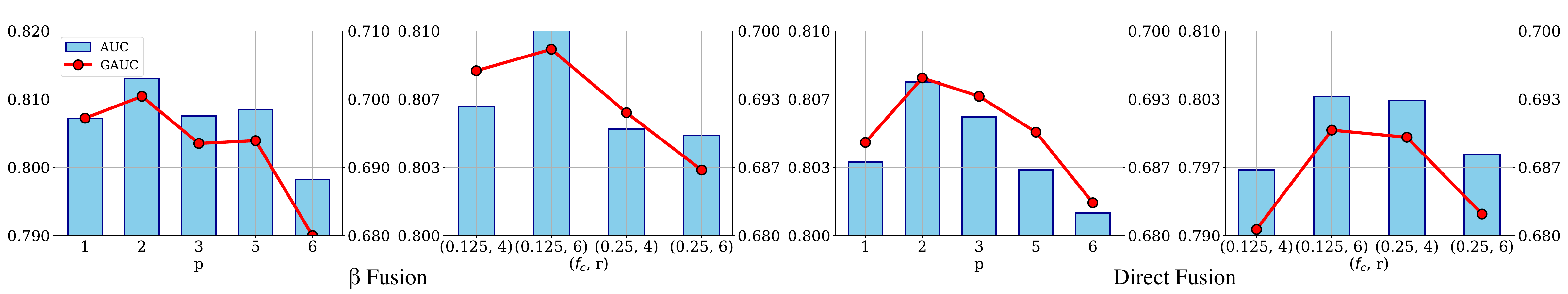}
     \caption{Comparison of truncation positions and filter parameters on FPEM's AUC/GAUC metrics for the Taobao dataset under various fusion strategies. $p$ is the truncation position, and $(f_c, r)$ represents the filter's cutoff frequency and order.}
    \label{fig:enter-label}
\end{figure*}

\textbf{FPEM Design Rationality Assessment}. 
In Table \ref{fft_design}, we explore different design strategies to enhance the performance of FPEM. Our experiments focus on two main aspects. First, we examine how to separate different frequencies, trying both truncation and Butterworth filtering methods. The filtering approach merges information from different frequency bands, while the truncation approach uses only the complete information from a some frequency band. The experiments show that merging different frequency bands does not bring positive gains to the module. Compared to the truncation method, the filtering method results in a 1.73\%/1.42\% decrease in AUC/GAUC on the Book dataset. Second, we explore how to merge these different frequency features, comparing direct fusion with $\beta$ fusion. High and mid-frequency features often contain users' exploratory shopping behavior, which is noise for periodic features. $\beta$ fusion uses target attention mechanism to suppresses some of this noise. The experiments show that this fusion method leads to a 0.68\%/0.72\% improvement in AUC/GAUC on the Taobao dataset compared to direct addition.

Table \ref{ablation_no_grad} shows the ablation experiments for the Frequency-Aware Module. It can be found that the performance of the model on both Book and Taobao datasets is significantly degraded without adding the information of items and users. In addition to this, the performance of the model is optimized if the added features of items and users are not back-propagated by the gradient at that stage. This is because the backpropagation of the gradient of the same feature in different modules will cause interference, which is not conducive to the effective learning of features.

\begin{table}[]
\centering
\caption{Ablation study for Frequency-Aware Module.}
\renewcommand{\arraystretch}{1.5}
\setlength\tabcolsep{4.2pt}
\resizebox{\columnwidth}{!}{
\begin{tabular}{ccc|cccc}
    \toprule
  \multirow{2}{*}{\makecell{\textbf{Pay\_Len}\\\textbf{Embeddings}}}  & \multicolumn{2}{c}{\makecell{\textbf{Item+User Side}\\ \textbf{Info Embeddings}}} & \multicolumn{2}{c}{\textbf{Book Dataset}} & \multicolumn{2}{c}{\textbf{Taobao Dataset}} \\ \cmidrule(r){2-3} \cmidrule(r){4-5} \cmidrule(r){6-7}
 & with grad            & w/o grad         & \textbf{AUC}         & \textbf{GAUC}       & \textbf{AUC}        & \textbf{GAUC}      \\ \cmidrule(r){1-7}
\checkmark   & \ding{55}                &  \ding{55}             & 0.7498 & 0.7270 & 0.8039 & 0.6881     \\
\checkmark   & \checkmark               &   \ding{55}            & 0.7540 & 0.7318 & 0.8077 & 0.6918     \\
\checkmark   &  \ding{55}               & \checkmark             & 0.7583 & 0.7341 & 0.8130 & 0.7004    \\ \bottomrule
\end{tabular}
}
\label{ablation_no_grad}
\end{table}

In Figure \ref{c and fc}, we further validate the conclusions regarding the selection of frequency band features. The first graph shows that when the truncation position is set to 5, the module’s AUC/GAUC reaches its peak and then begins to decline. In the second graph, since the input sequence length for the FPEM module on the Book dataset is 26, the corresponding truncation position is 3.25 when the cutoff frequency $f_c$ is 0.125 and 6.5 when the $f_c$ is 0.25. And the higher the filter order $r$, the closer the filter function approximates a step function, and the experimental results confirm this observation. Specifically, the AUC/GAUC metrics are similar when c=3 in the first graph and when $f_c$=0.125, $r$=6 in the second graph, but there is some difference when $f_c$=0.125, $r$=4. Moreover, the experiments also show that scaling the original spectral information has negative effects; the truncation method, which retains the complete spectral information, outperforms the filter method. 
In Figure \ref{fig:enter-label}, the experiments regarding the selection of frequency band features on the Taobao dataset are further substantiated. The first graph demonstrates that when the truncation position is set to 2, the module's AUC and GAUC reach their maximum values and subsequently decline. This finding underscores the presence of substantial noise within the high-frequency segment, suggesting that retaining excessive noise features might compromise the predictive capabilities of the model.
Similarly, the experimental findings depicted in the second graph are consistent with the observations made with the Book dataset. As the truncation position escalates, the AUC and GAUC values initially ascend to a peak before experiencing a downturn. This pattern implies that judiciously eliminating high-frequency noise features can significantly improve the accuracy and stability of the model.
Both sets of experimental outcomes reinforce the analysis's validity, emphasizing the critical role of choosing an optimal truncation position for enhancing model performance when working with spectral data. This approach ensures that the model effectively mitigates noise, thereby boosting its overall performance.

\subsection{Online A/B Test (RQ3)}
To assess the online performance of FIM, we conduct a two-week A/B test on Kuaishou’s local-life service platform. In the first week, we evaluate our Multi-View Search Strategy (MSS). In the second week, the Frequency-Domain Perception and Evolution Module (FPEM) is incorporated. GMV, the number of paid orders, and the number of buyers are utilized as metrics for online evaluation. As shown in Table \ref{online}, MSS boosts GMV, paid orders, and buyers, demonstrating its strong ability to model users' interests. Incorporating FPEM in the second week allows the model to capture users' periodic features, further increasing paid orders and buyers while maintaining positive GMV gains.

\begin{table}[]
\centering
\caption{Business gains from the phased online A/B testing of FIM on Kuaishou’s local-life service platform. The first week uses only the MSS module, while the second week’s experiment is based on the results of the first week.}
\renewcommand{\arraystretch}{1.5}
\resizebox{0.9\columnwidth}{!}{%
\begin{tabular}{c|ccc}
\toprule
\textbf{Method}        & \textbf{GMV}      & \textbf{Order\_Num} & \textbf{Buyer\_Num} \\ \cmidrule(r){1-4}
Base vs MSS     & +8.462\% & +7.122\%   & +6.010\%   \\
MSS vs MSS+FPEM & +3.293\% & +7.194\%   & +6.087\%   \\ \bottomrule
\end{tabular}%
}
\label{online}
\end{table}

To validate the effectiveness of our model in handling different frequency interests, we conducted an in-depth analysis of the experimental results through online A/B testing. The offline Kuaishou dataset contains user information spanning a month, while the online dataset includes the user's entire lifetime sequence. In the local life scenario of Kuaishou, it currently contains approximately two years of data.
Through this analysis, it was found in Table \ref{online2} that our Frequency-Aware Multi-View Interest Modeling (FIM) framework has significant gains in various in-depth areas compared to the baseline. Notably, the model brings significant benefits in both high-frequency consumption scenarios, such as dining, and low-frequency consumption scenarios, such as hotel \& travel.
The offline time and space complexity analysis of the Multi-View Search Strategy (MSS) and Frequency-domain Perception and Evolution Module (FPEM) is provided in Table \ref{aball}. As for the added system latency when deployed online, our model increases the time consumption by 11.245\%. However, despite this increase in time consumption, the significant improvements in recommendation accuracy and user satisfaction make the additional latency well worth it. These modules, MSS and FPEM, work together to enhance the model's ability to handle different frequency interests, leading to significant improvements in both high-frequency and low-frequency consumption scenarios. The combination of these strategies ensures that the model can effectively capture and adapt to the dynamic nature of user preferences, making it a robust solution for local-life service platforms.

\begin{table}[]
\centering
\caption{Business gains from the phased online A/B testing of FIM on Kuaishou’s local-life service platform. The first week uses only the MSS module, while the second week’s experiment is based on the results of the first week.}
\renewcommand{\arraystretch}{1.5}
\resizebox{0.85\columnwidth}{!}{%
\begin{tabular}{c|ccc}
\toprule
\textbf{Category} & \textbf{GMV Improv. (\%)} & \textbf{Order Improv. (\%)} \\
\hline
Dining & 9.77 & 6.90 \\
General & 8.51 & 10.35 \\
Hotel \& Travel & 8.28 & 6.81 \\ \bottomrule
\end{tabular}%
}
\label{online2}
\end{table}
\section{Conclusion}
In this paper, to model users' periodic behavior in local-life service scenarios, we propose an innovative frequency-aware multi-view interest modeling network, named FIM. Specifically, we design a multi-view search strategy to precisely and efficiently capture users' interest preferences. Additionally, we introduce a plug-and-play frequency perception and evolution module to capture changing periodic preferences.
Extensive experiments on two public datasets and the Kuaishou industrial dataset, along with online A/B tests, empirically demonstrate the superiority of our model. 
For future research, we plan to incorporate factors such as marketing signals that influence user behavior to dynamically capture changes in their periodic interests.

\bibliographystyle{ACM-Reference-Format}
\balance
\bibliography{sample-base}

\end{document}